\documentclass[aps,superscriptaddress,prl,groupaddress,twocolumn,floatfix]{revtex4-2}
\usepackage{units}
\usepackage{amsmath}
\usepackage{amssymb}
\usepackage{graphicx}
\usepackage{bm}
\usepackage{multirow, color,relsize,ulem,microtype}

\begin{document}
\title{Nonlinear Photonic Tripartite Phase}

\author{Xiangrui Hou}
\altaffiliation[]{Contributed equally}
\author{Fangyu Wang}
\altaffiliation[]{Contributed equally}
\author{Zhaoxin Wu}
\author{Shuming Zhang}
\affiliation{School of Physics and Zhejiang Key Laboratory of Micro-nano Quantum Chips and Quantum Control, Zhejiang University, Hangzhou 310058, Zhejiang Province, China}
\author{Shan-Zhong Li}
\affiliation{Key Laboratory of Atomic and Subatomic Structure and Quantum Control (Ministry of Education),
Guangdong Basic Research Center of Excellence for Structure and Fundamental Interactions of Matter,
School of Physics, South China Normal University, Guangzhou 510006, China}
\author{Lei Ying}
\author{Haiqing Lin}
\affiliation{School of Physics and Zhejiang Key Laboratory of Micro-nano Quantum Chips and Quantum Control, Zhejiang University, Hangzhou 310058, Zhejiang Province, China}
\author{Baile Zhang}
\affiliation{Division of Physics and Applied Physics, School of Physical and Mathematical Sciences, Centre for Disruptive Photonic Technologies, Nanyang Technological University, Singapore, Singapore}
\author{Zhi Li}
\email{lizphys@m.scnu.edu.cn}
\author{Shi-Liang Zhu}
\email{slzhu@scnu.edu.cn}
\affiliation{Key Laboratory of Atomic and Subatomic Structure and Quantum Control (Ministry of Education),
Guangdong Basic Research Center of Excellence for Structure and Fundamental Interactions of Matter,
School of Physics, South China Normal University, Guangzhou 510006, China}
\author{Zhaoju Yang}
 \email{zhaojuyang@zju.edu.cn}
\affiliation{School of Physics and Zhejiang Key Laboratory of Micro-nano Quantum Chips and Quantum Control, Zhejiang University, Hangzhou 310058, Zhejiang Province, China}

\begin{abstract}
Anderson localization is usually understood as a transition between extended and localized phases, with criticality confined to a single mobility edge. Recent advances predict that quasiperiodic systems can instead host a finite critical window bounded by mobility edges, in which localized, critical and extended states coexist. Yet both the experimental realization of this regime and whether interactions can provide controlled access to it remain unknown. Here, we realize such a tripartite phase in a nonlinear quasiperiodic photonic lattice and show that Kerr nonlinearity, acting as an effective interaction, enables state-selective access to the critical window. By tracking wavepacket dynamics, we distinguish localized, critical and extended transport regimes and uncover a state-selective response: rather than simply reinforcing localization through self-trapping, weak nonlinearity drives low-energy localized states into the critical window, whereas stronger nonlinearity restores localization. By contrast, critical, extended and high-energy localized states evolve monotonically towards self-trapped behaviour. Our results reveal a state-selective mechanism by which interactions provide controlled access to a pre-existing critical window in quasiperiodic systems.


\end{abstract}

\maketitle
Anderson localization is a central paradigm of wave transport in disordered media~\cite{pwanderson1958pr, abrahams1979prl,palee1985rmp,fevers2008rmp}, with manifestations in electronic~\cite{palee1985rmp}, photonic~\cite{Schwartz2007TransportLatticesb, Lahini2008AndersonLatticesb, Lahini2009ObservationLattices}, and atomic~\cite{Billy2008DirectDisorder, Roati2008AndersonCondensate} systems. In the conventional mobility-edge picture, extended and localized states are separated by a single critical energy~\cite{fevers2008rmp}, where wavefunctions acquire scale-invariant fluctuations and multifractal structure~\cite{admirlin2006prl,hyao2019prl}. In this sense, criticality appears only at a finely tuned spectral boundary: an infinitesimal shift in energy restores either transport or localization~\cite{sganeshan2015prl, Li2015Many-BodyEdge,hluschen2018prl,faan2018prx,ywang2020prl,falexan2021prl,ywang2022prl,jgao2025sb,hthu2025prl, Chang2025ObservationPhasesb}. Recent theoretical advances have pointed to a broader possibility in quasiperiodic lattices~\cite{aaubry1980aips,rketzmerick1997prl,fliu2015prb,jwang2016prb,txiao2021sb,Wang2021Many-BodyNonthermal,yczhang2022prb,hli2023npj}, where a single mobility edge can split into two boundaries enclosing a finite interval of critical states~\cite{xdeng2019prl,ywang2022prb,mgoncalves2023prl,xczhou2023prl,szli2025scpma,Zhou2026TheResults}. The resulting spectrum hosts localized, critical and extended states within a single tripartite phase, extending multifractal criticality from an isolated energy to a finite spectral interval. What remains missing, however, is an experimental realization of this tripartite phase together with a controllable route to dynamically access its critical window.

A central open question is whether interactions can provide such control, beyond conventional tuning by disorder strength or energy~\cite{Belitz1994TheTransition, Segev2013AndersonLight, Abanin2019Colloquium:Entanglement}. Previous studies have largely focused on whether interactions reinforce localization ~\cite{Schwartz2007TransportLatticesb, Lahini2008AndersonLatticesb, Schreiber2015ObservationLattice} or promote delocalization~\cite{Pikovsky2008DestructionNonlinearity, Flach2009UniversalSystems, Vakulchyk2019WaveWalks, SeeToh2022Many-bodyGas, Cao2022Interaction-drivenGas}. Much less is known about how interactions act in spectra that already contain a finite critical window. In particular, it remains unclear whether effective interactions can drive initially localized states into that window or instead push them farther away, and whether critical transport can persist over a finite interaction range rather than appearing only at isolated transition points. Resolving these questions is essential for establishing interactions as a controlled route for state-selective access to critical transport in quasiperiodic systems.

Here, we experimentally realize a nonlinear quasiperiodic photonic lattice hosting a tripartite phase in which localized, critical and extended states coexist, and show that effective interactions~\cite{Maczewsky2020Nonlinearity-inducedInsulator, Mukherjee2020ObservationBandgapc, Jurgensen2021QuantizedPumping, Jurgensen2023QuantizedSolitons} open dynamical access to the critical window. By directly tracking wavepacket dynamics, we resolve distinct transport signatures of the three spectral sectors and uncover a state-selective interaction response: weak nonlinearity drives low-energy localized states into the critical window before stronger nonlinearity restores localization through self-trapping, whereas critical, extended and high-energy localized states evolve monotonically towards self-trapping. Together, these observations establish a nonlinear photonic tripartite phase and reveal a state-selective mechanism for accessing the critical window: interaction-induced spectral flow, together with mobility-edge asymmetry, determines whether a state is driven into the critical window or further into localization.

\begin{figure*}[t]
\includegraphics[width=1\linewidth]{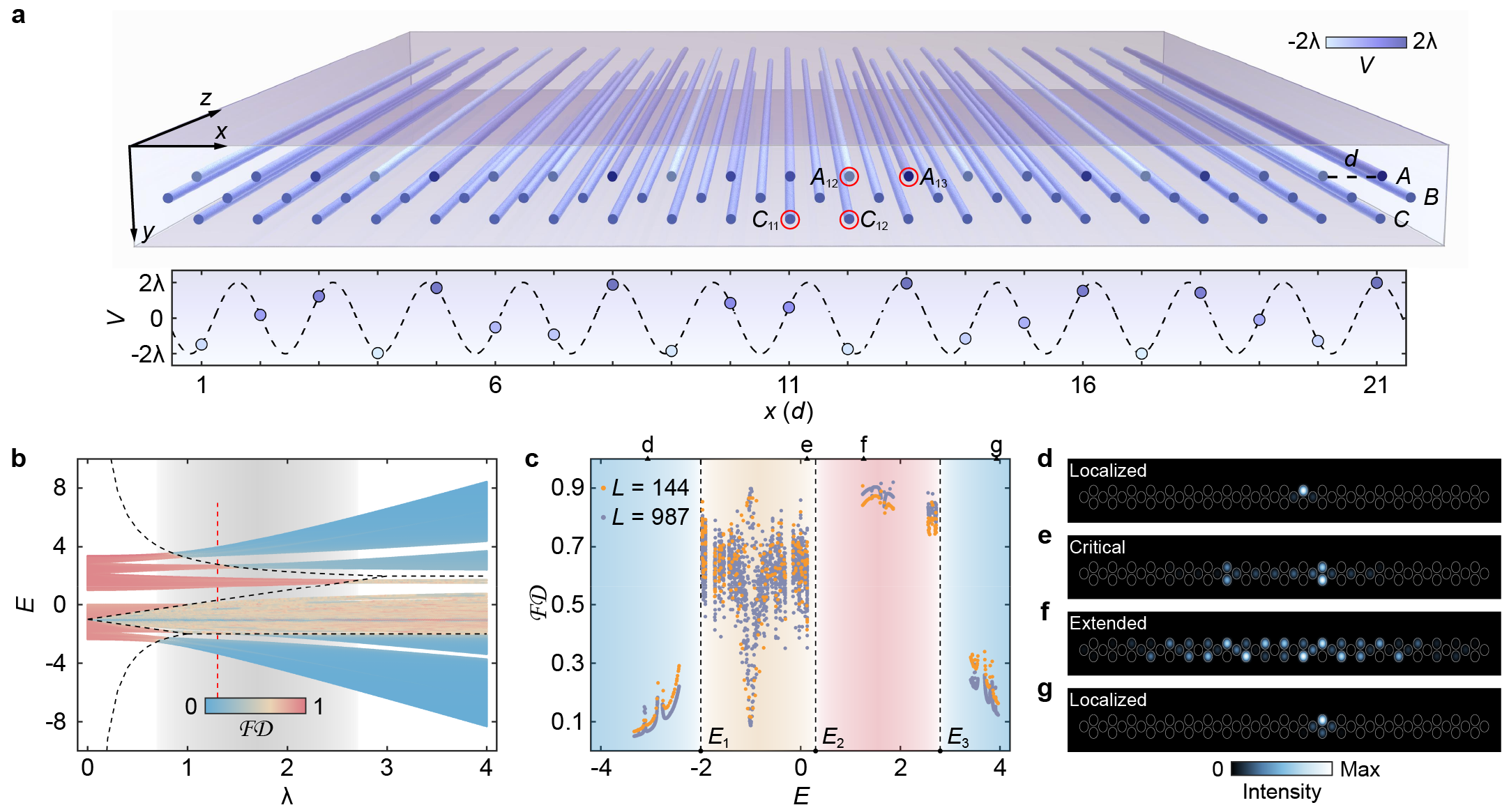}
\caption{\textbf{Experimental realization of a tripartite phase in a quasiperiodic diamond lattice.} \textbf{a,} Schematic of the photonic waveguide array, where the quasiperiodic potential applied to sublattice $A$ is depicted by the purple color gradient. The superimposed curve illustrates the on-site potential energy distribution. The red circles mark the selective excitation sites targeting the low-energy localized~($A_{12}$), critical~($C_{11}$), extended~($C_{12}$), and high-energy localized~($A_{13}$) states. \textbf{b,} Phase diagram for a system size of $L=987$, with black dashed lines indicating mobility edges. \textbf{c,} Spectrum of the fractal dimension $\mathcal{FD}$. The dependence of $\mathcal{FD}$ on energy $E$ is plotted for a fixed disorder strength $\lambda=1.3$ (vertical red dashed line in \textbf{b}). Data points compare finite-size scaling results for $L=144$ (orange) and $L=987$ (slate blue). The background shading in \textbf{b} and \textbf{c} distinguishes the localized (blue), critical (yellow), and extended (red) sectors. \textbf{d-g,} Representative eigenstate profiles for the experimental lattice size ($L=21$), corresponding to the energies marked by black triangles in \textbf{c}. All parameters are normalized to the coupling strength $J=t=1$, with lattice constant $d$.}
\label{fig:1}
\end{figure*}

\begin{figure*}[t]
\includegraphics[width=1\linewidth]{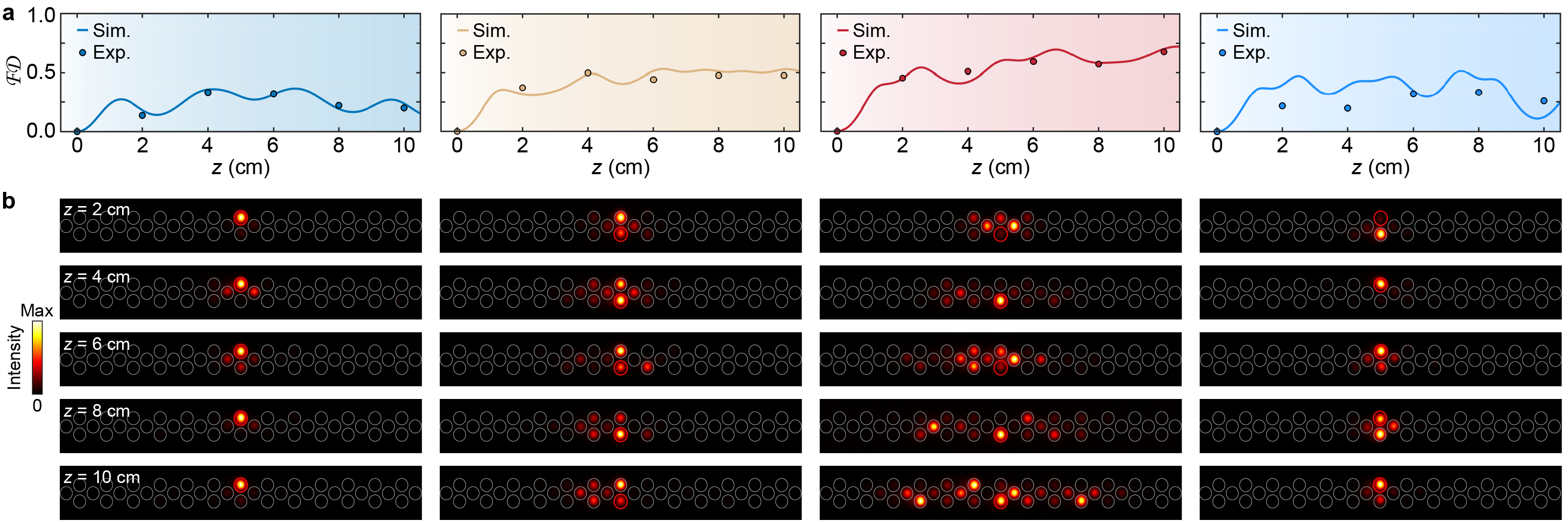}
\caption{\textbf{Dynamical signatures of the tripartite phase. a,} Evolution of the fractal dimension $\mathcal{FD}$ as a function of propagation distance $z$. The experimental measurements (dots) are consistent with the numerical simulations (solid lines). \textbf{b,} Spatial intensity distributions measured at propagation lengths of $z=$2, 4, 6, 8, and 10 cm. The columns illustrate three distinct transport regimes: the localized state remains spatially confined (1st and 4th column), the critical state exhibits anomalous expansion (2nd column), and the extended state undergoes ballistic spreading (3rd column). Lattice parameters: system size $L=21$, coupling strength $J=t=0.5\,\text{cm}^{-1}$, and disorder strength $\lambda=1.3J$.}
\label{fig:2}
\end{figure*}

\textbf{Model Hamiltonian}. To experimentally probe the coexisting states, we harness the quasiperiodic diamond photonic lattice consisting of coupled optical waveguides (Fig.~\ref{fig:1}a). As a paradigmatic flat-band system, this diamond model offers a distinct advantage: its spectral structure allows for the analytical prediction of the mobility edges via global theory~\cite{aavila2015am}, providing a useful benchmark for the experiments. The system dynamics are governed by the Hamiltonian:

\begin{equation}
H=\sum_{j=1}^{L}\Psi_j^\dagger h_0 \Psi_j
+\sum_{j=1}^{L-1}\left(\Psi_j^\dagger h_1 \Psi_{j+1}+\text {h.c.}\right).
\label{eq:1}
\end{equation}
Here $\Psi_j=[a_j, b_j, c_j ]^{\mathrm T}$, $h_0=[V(j),J,t;J,0,J;t,J,0]$ and $h_1=[0,J,0;0,0,0;0,J,0]$. $a_{j}^{\dagger} (a_{j})$, $b_{j}^{\dagger} (b_{j})$ and $c_{j}^{\dagger} (c_{j})$ denote the creation (annihilation) operators acting on sublattices $A$, $B$, and $C$ in the $j$-th unit cell, respectively. The parameters $J$ and $t$ represent the nearest-neighbor (between $B$ and $A/C$) and next-nearest-neighbor couplings (between $A$ and $C$), respectively. The quasiperiodic modulation $V(j)=2\lambda \cos(2\pi\alpha j + \theta)$ is exclusively applied to the $A$ sublattices, where $\lambda$, $\alpha$, and $\theta$ correspond to the potential strength, irrational number, and phase offset. $L$ denotes the total number of unit cells. Without loss of generality, we set $\alpha=(\sqrt5-1)/2$, $\theta=0$ and fix $J=t=1$ as the energy unit. Experimentally, we realize the model described by Eqn.~\eqref{eq:1} using femtosecond-laser-written waveguide arrays \cite{rechtsman2013n}. In the paraxial approximation, the propagation of light through these evanescently coupled waveguides is governed by a Schr\"{o}dinger-like equation, effectively mapping the spatial propagation coordinate $z$ to the temporal evolution $t$ in a quantum system~\cite{rechtsman2013n, Biesenthal2022FractalInsulators, ysun2024prl}.

To quantitatively characterize the localization, we use the Fractal Dimension ($\mathcal{FD}$)~\cite{fevers2008rmp}, a standard order parameter in the study of Anderson transitions. It is defined as $\mathcal{FD}=-\lim_{L\rightarrow\infty}\frac{\ln(\text{IPR})}{\ln(3L)}$,
where the Inverse Participation Ratio (IPR) is given by $\text{IPR}=\Sigma_{j,u}|\phi_{j,u}|^4$, with $\phi_{j,u}$ representing the normalized amplitude distribution on sublattice $u \in {A, B, C}$ in the $j$-th unit cell. In the thermodynamic limit, $\mathcal{FD}$ approaches 1 for extended states and 0 for localized states, while lying between 0 and 1 for critical states.

Figure~\ref{fig:1}b maps the $\mathcal{FD}$ spectrum as a function of the potential strength $\lambda$. The boundaries separating the extended, critical, and localized states are marked by black dashed lines, corresponding to the mobility edges that can be analytically derived~\cite{aavila2015am, szli2025scpma}. To distinguish these states, we perform a finite-size scaling analysis with fixed $\lambda=1.3$, corresponding to the red dashed line in Fig.~\ref{fig:1}b. The result is presented in Fig.~\ref{fig:1}c and Supplementary Information, section I-A. The predicted mobility edges are located at $E_1=-2.0$, $E_2=0.2$, and $E_3=2.8$, which partition the energy spectrum into three sectors of the tripartite phase. For states with energies in the localized sectors ($E<E_1$ or $E>E_3$), the $\mathcal{FD}$ decays monotonically with increasing system size of $L$, tending towards zero. Conversely, for states in the extended sector ($E_2<E<E_3$), the $\mathcal{FD}$ exhibits an upward trend, asymptotically approaching unity. Notably, within the critical window ($E_1<E<E_2$), the $\mathcal{FD}$ manifests a robust scale-invariant behaviour \cite{xczhou2023prl, hli2023npj}, maintaining a constant fractional value regardless of system size, which is the hallmark of multifractality. These distinct scaling behaviours provide a rigorous theoretical basis for identifying the different states in our finite-sized experimental system. The typical spatial distributions of eigenstates representative of the tripartite phase are visualized in Fig.~\ref{fig:1}d-g. This spectral asymmetry becomes important below, because under positive nonlinear detuning, the critical window is adjacent only to the low-energy localized sector.

\textbf{Experimental implementation}. To experimentally access the coexisting states, we fabricate the quasiperiodic diamond lattice using the femtosecond laser writing method~\cite{rechtsman2013n} (Supplementary Information II-A). We explicitly tailor the lattice geometry to satisfy the condition $J=t$ by meticulously adjusting the waveguide spacings. Simultaneously, the quasiperiodic on-site potential is imprinted by modulating the laser writing velocity, allowing us to precisely detune the refractive index of individual waveguides~\cite{Hou2026QuantumLight}. In the experiments, the disorder strength is set to be $\lambda = 1.3J$, where the system supports a tripartite phase with localized, critical, and extended states.

To distinguish these states, we develop a site-selective excitation protocol. While exciting the eigenstates in a continuum is challenging, single-site excitations allow us to prepare initial wavepackets with high overlap fidelity to specific target states—localized, extended, or critical. By monitoring the spatial spreading of the wavefunction along the propagation axis $z$ (acting as time), we can extract the evolution of the fractal dimension $\mathcal{FD}$. This dynamical observable serves as a robust order parameter: the wavepacket density profiles and their associated $\mathcal{FD}$ converge to the characteristic values of the underlying state (Supplementary Information I-B, D), providing a direct signature of the modeled Hamiltonian.

\begin{figure*}[t]
\includegraphics[width=0.8\linewidth]{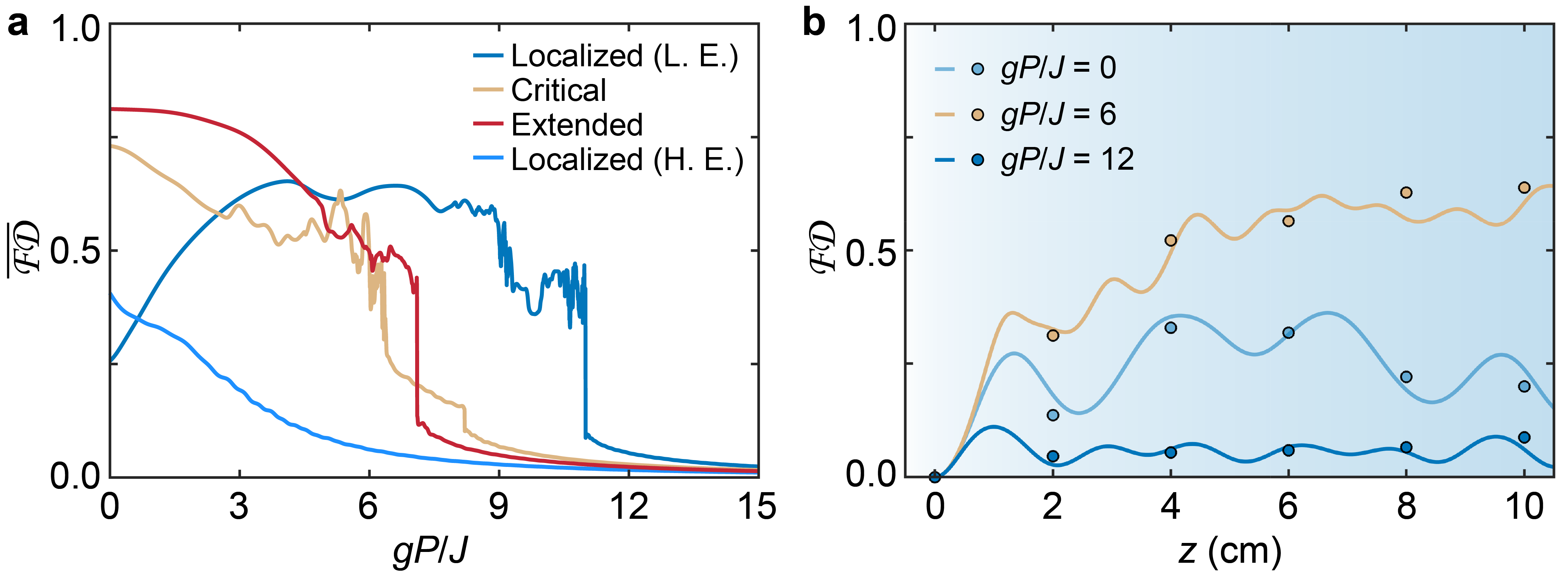}
\caption{\textbf{Interaction response in the tripartite phase.} \textbf{a,} The time-averaged fractal dimension $\overline{\mathcal{FD}}$ as a function of interaction strength, for the 
low-energy (L. E.) localized~($A_{12}$), critical~($C_{11}$), extended~($C_{12}$), and high-energy (H. E.) localized~($A_{13}$) states. The value is integrated over the propagation window $z \in [20, 50] $cm. \textbf{b,} Dynamical evolution of the $\mathcal{FD}$ for the three interaction strengths. Experimental measurements (dots) show excellent agreement with numerical simulations (solid lines), confirming the non-monotonic transition. All parameters are consistent with Fig.~\ref{fig:2}. }
\label{fig:3}
\end{figure*}

In the following, we experimentally investigate the dynamics by launching a probe beam ($\lambda_0$ =1030 nm) into sites specifically chosen to resonate with the distinct spectral sectors: site $A$ of the 12th cell, site $C$ of the 11th cell, site $C$ of the 12th cell, and site $A$ of the 13th cell, for exciting the low-energy localized, critical, extended, and high-energy localized states, respectively. Figure~\ref{fig:2}b visualizes the spatial intensity evolution captured at the output facet for propagation lengths of $z$=2, 4, 6, 8, and 10 cm. The white circles represent the waveguide lattice, and the red circles indicate the position of the excited waveguides.

The transport dynamics exhibit striking contrasts. The localized state (1st and 4th column) remains spatially confined near the injection site with a frozen intensity profile, a hallmark of Anderson localization. Conversely, the extended state (3rd column) undergoes rapid, ballistic expansion across the lattice. The critical state instead exhibits anomalous spreading: the wavepacket neither remains frozen nor expands ballistically, but undergoes extended yet non-ergodic dynamics within a finite spatial region. Our experimental measurements are consistent with the numerical simulations (see detailed results in Supplementary Information I-C).

To quantify these observations, we extract the measured $\mathcal{FD}$ evolution (dots) and compare it with numerical simulations (solid curves in Fig.~\ref{fig:2}a). As we can see, the experimental data agree well with the numerical curves. The low-energy (high-energy) localized wavepacket maintains a low dimensionality, fluctuating around $\mathcal{FD}\sim 0.25 \ (0.4)$. The extended state rapidly spreads across the lattice, with $\mathcal{FD}$ approaching the value of 0.75. Notably, the critical state sustains an intermediate dimensionality of $\mathcal{FD}\sim 0.50$ throughout the evolution, distinct from both limits. The agreement between experiment and theory supports the mobility-edge-bounded tripartite phase inferred from the theoretical model. Notably, the observed $\overline{\mathcal{FD}}$ values deviate from the idealized thermodynamic limits—$0$ for localized and $1$ for extended states—owing to finite-size constraints (see Supplementary Information I-A, B for scaling analysis).

\begin{figure*}[t]
\includegraphics[width=1\linewidth]{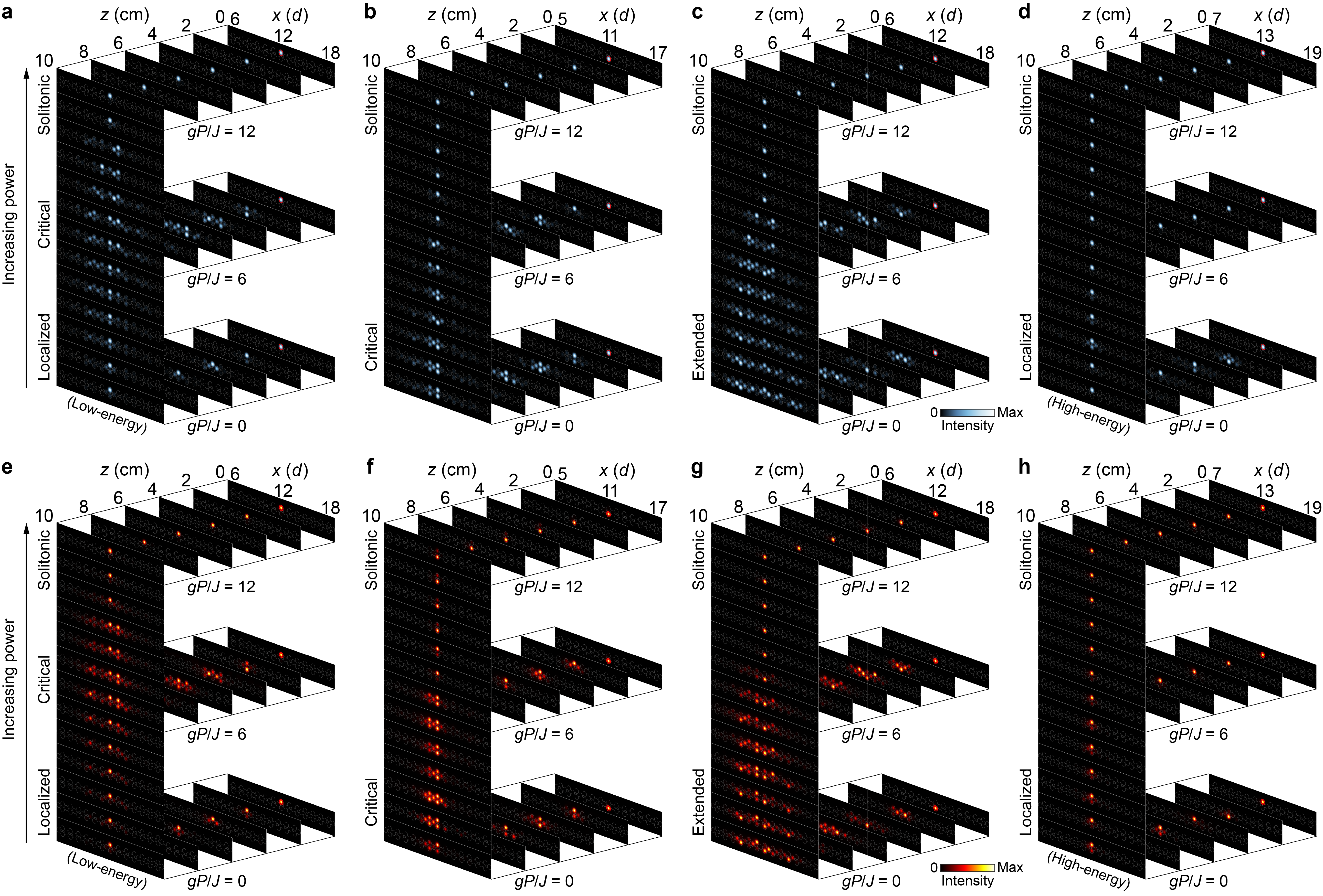}
\caption{\textbf{State-selective nonlinear transport and re-entrant localization.} Numerical (\textbf{a-d}) and experimental (\textbf{e-h}) spatial intensity distributions for the low-energy localized state (\textbf{a, e}), critical state (\textbf{b, f}), extended state (\textbf{c, g}), and high-energy localized state (\textbf{d, h}). Each panel displays wavepacket profiles stacked vertically in ascending order of input power $gP/J$. A striking contrast is observed: the low-energy localized state (\textbf{a, e}) exhibits a unique non-monotonic anomalous transition (localization $\to$ critical transport $\to$ soliton), whereas the critical, extended, and high-energy localized states (\textbf{b-d, f-h}) undergo monotonic self-trapping, directly collapsing into the soliton regime. This comparison confirms that the interaction-induced delocalization is observed only for the low-energy localized sector. }
\label{fig:4}
\end{figure*}

\textbf{Interaction-enabled criticality}. Having established the static spectral topology, we next ask how interactions reshape the localized, critical, and extended sectors. One might expect interactions simply to reinforce localization through self-trapping. Instead, we find a strikingly non-monotonic evolution: weak interactions drive the initially low-energy localized states into the critical window, whereas stronger interactions restore localization through self-trapping.

In our photonic platform, this interaction effect is implemented at the mean-field level through an intensity-dependent on-site Kerr nonlinearity in the nonlinear lattice. The resulting wave dynamics are governed by the discrete nonlinear Schr\"{o}dinger equation
~\cite{Maczewsky2020Nonlinearity-inducedInsulator, Mukherjee2020ObservationBandgapc, Jurgensen2021QuantizedPumping, Jurgensen2023QuantizedSolitons}:

\begin{equation}
    -i\partial_{z}\psi_j= H\psi_j+g|\psi_j|^2\psi_j,
\end{equation}
where $\psi_j$ represents the wavefunction (the envelope of the electric field) in the $j$th waveguide. $g$ describes the strength of nonlinearity and is positive (negative) for a focusing (defocusing) Kerr nonlinearity. 
To compare the experimental and numerical results hereafter, we define the input power $P=\Sigma_{j}|\psi_{j}(z=0)|^2$, which refers to the strength of nonlinearity as a dimensionless quantity of $gP/J$. Note that the nonlinear Schr\"{o}dinger equation with $g>0$ is equivalent to an attractive Gross-Pitaevskii equation describing Bose-Einstein condensation. This setup allows us to explore the interplay between the quasiperiodic potential and interaction, revealing a non-monotonic transition from localized to critical transport and ultimately to self-trapping. 

To quantitatively map this interaction-driven transition, we employ the time-averaged fractal dimension as a dynamical order parameter $\overline{\mathcal{FD}}=\frac{1}{z_f-z_i}\int_{z_i}^{z_f}\mathcal{FD}(z)dz$, where $\mathcal{FD}(z)$ captures the instantaneous dimensionality of the wavepacket during the propagation. Figure~\ref{fig:3}a presents the tight-binding numerical results of the dynamical order parameter $\overline{\mathcal{FD}}$ as a function of the strength of nonlinearity $gP/J$, for the low-energy localized, critical, extended, and high-energy localized states. The results obtained over alternative integration windows are provided in Supplementary Information I-E, and they exhibit agreement with each other. 

Surprisingly, the interaction-dependent phase diagram shows that, starting from the linear regime ($gP/J\rightarrow0$), the system resides first in the low-energy localized sector with about $\overline{\mathcal{FD}}\sim 0.25$. As the nonlinearity increases, we observe a distinct delocalization transition: the wavefunction expands into the critical window, characterized by a saturation plateau of $\overline{\mathcal{FD}}>0.5$ within the window of about $2<gP/J<9$. This indicates that the weak interaction effectively melts the localized state into the critical window of the tripartite phase. Upon further increasing the interaction strength ($gP/J>9$), the system undergoes re-entrant localization, and eventually collapses into a self-trapped soliton ($\overline{\mathcal{FD}}\rightarrow0.1$) for $gP/J>11$, where the interaction-induced potential effectively decouples the waveguide and the wavefunction is trapped. The same trend persists across the low-energy localized sector (Supplementary Information I-E). In stark contrast, with the launch of critical, extended, and high-energy localized states, we observe that the dimensionalities exhibit monotonic decreasing behaviour, which confirms the conventional wisdom that Kerr nonlinearity enhances localization through self-trapping.

The underlying mechanism is governed by state-selective nonlinear energy renormalization and the asymmetry of the mobility-edge structure. As detailed in our perturbation analysis (Supplementary Information I-F), the Kerr nonlinearity acts as a positive diagonal potential, $V_{\text{nl}}=g|\psi|^2\;(g>0)$, inducing a first-order energy shift proportional to the mode’s $\mathrm{IPR}$, $\Delta E \propto gP\cdot\mathrm{IPR}$. Consequently, localized states, possessing maximal $\mathrm{IPR}$, experience a rapid and positive deterministic spectral drift, while extended and critical states remain energetically pinned due to their comparatively small $\mathrm{IPR}$.

However, the ultimate fate of the state is dictated by the direction of the spectral flow relative to the mobility edges. For the low-energy localized state positioned below the lower mobility edge, the positive detuning drives the eigenenergy upwards across the mobility edge and into the critical window, triggering a resonant hybridization with the critical states. In stark contrast, for a high-energy localized state positioned above the upper mobility edge, the same positive drift propels the energy further away from the band continuum, reinforcing localization and leading directly to self-trapping. The above discussion is corroborated by the eigenmode analysis (Supplementary Information I-F). We therefore identify a general mechanism for state-selective access to the critical window, set by the IPR-weighted nonlinear energy renormalization and the asymmetry of mobility edges.

With experiments, we visualize this non-monotonic transition and present the results in Fig.~\ref{fig:4}e. The spatial intensity profiles reveal three distinct transport sectors governed by the interplay between disorder and the strength of nonlinearity. At zero nonlinearity ($gP/J=0$), the wavepacket remains frozen in the localized sector. As the input power increases to an intermediate level ($gP/J=6$), the wavepacket undergoes an anomalous expansion, evolving into a critical window characterized by a multifractal feature of $\mathcal{FD} \sim 0.6$. Finally, at high power ($gP/J=12$), the system enters a re-entrant localization regime, where the wavepacket collapses into a tight soliton. The experimental results (Fig.~\ref{fig:4}e) are in excellent agreement with the numerical simulations (Fig.~\ref{fig:4}a). The experimental $\mathcal{FD}$ evolutions exhibit excellent agreement with numerical simulations (Fig.~\ref{fig:3}b). Specifically, the light-blue, yellow, and dark-blue curves delineate the distinct dynamical signatures of the localized, critical, and solitonic (self-trapping) regimes, respectively, capturing the non-monotonic transition driven by the varying interaction. These measurements confirm that nonlinearity can drive the frozen low-energy localized states into the critical window before self-trapping dominates at stronger nonlinearity.

Fundamentally, this interaction-induced delocalization is unique to the low-energy localized states positioned near the mobility edge. As shown in Fig.~\ref{fig:4}f-h, our experimental measurements show that starting from critical, extended, or high-energy localized states, the introduction of nonlinearity monotonically drives the system directly into a soliton state, in the absence of delocalization. The experimental measurements verify the simulated results of $\overline{\mathcal{FD}}$ (Fig.~\ref{fig:3}a) that, unlike the non-monotonic thawing behaviour observed in the low-energy localized state, the fractal dimensions of the critical, extended, and high-energy localized states exhibit a monotonic decay trend, indicating a direct collapse into the self-trapped soliton regime. 

In conclusion, we have experimentally realized a nonlinear photonic tripartite phase and shown that interactions can open dynamical access to a pre-existing critical window. Rather than simply reinforcing localization through self-trapping, weak nonlinearity drives low-energy localized states into the critical window, whereas stronger nonlinearity restores localization; by contrast, critical, extended and high-energy localized states evolve monotonically towards self-trapping. These results reveal a state-selective mechanism for accessing the critical window: interaction-induced, IPR-weighted spectral flow, together with mobility-edge asymmetry, determines whether a localized state is driven into critical transport or further into localization. Our work establishes a controlled photonic platform~\cite{Ozawa2019TopologicalPhotonics, Goblot2020EmergenceChains, Xia2021NonlinearStatesb} for exploring critical dynamics in quasiperiodic localization landscapes, with clear opportunities to extend these ideas to interacting~\cite{Rispoli2019QuantumTransition, Dong2024QuantumAtoms, Karamlou2024ProbingLattice, Yu2024ObservingQuasicrystal, Andersen2025ThermalizationSimulatorb, Guo2025ObservationLocalization} (Supplementary Information I-H) and dissipative~\cite{Zhao2025UniversalGases, Fang2025ProbingArrays} wave systems.

~\\

\begin{acknowledgments}
This research is supported by the National Key R\&D Program of China (Grant No. 2023YFA1406703, 2022YFA1404203, 2022YFA1405300), Innovation Program for Quantum Science and Technology (Grant No. 2021ZD0301705), Guangdong Provincial Quantum Science Strategic Initiative (Grant No. GDZX2304002), Zhejiang Provincial Natural Science Foundation of China (Grant No. LR23A040003), and Fundamental Research Funds for the Central Universities (Grant No. 226-2025-00124). BZ acknowledges support from the Singapore National Research Foundation Competitive Research Program under Grant No. NRF-CRP23-2019-0007 and the Singapore Ministry of Education Academic Research Fund Tier 2 under Grant No. MOE-T2EP50123-0007.
\end{acknowledgments}

%

\end{thebibliography}

\end{document}